\documentclass[12pt]{article}
\usepackage{amsmath,amssymb,mathrsfs,url}
\usepackage{cite}
\usepackage{authblk}

\title{Wave mechanics for gravity with point-particles}

\author[*]{Christian Maes}
\author[*]{Kasper Meerts}
\author[*,$\dagger$]{Ward Struyve}
\affil[*]{Instituut voor Theoretische Fysica, KU Leuven}
\affil[$\dagger$]{Centrum voor Logica en Filosofie van de Wetenschappen, KU 
Leuven}
\date{}

\def\de{\delta}

\def\na{\nabla}

\def\ka{\kappa}

\def\pa{\partial}

\def\ka{\kappa}
\def\ii{\textrm i}
\def\ee{\textrm e}

\newcommand{\be}{\begin{equation}}
\newcommand{\en}{\end{equation}}
\newcommand{\bi}{\begin{itemize}}
\newcommand{\ei}{\end{itemize}}

\begin{document}
\maketitle

\begin{abstract}
\noindent
We consider non-relativistic point-particles coupled to Einstein gravity and 
their canonical quantization. From the resulting Wheeler-DeWitt wave equation we determine a quantum version of geometrodynamics, where the coupled evolution of particle positions and 3-metric is guided by the wave function.  We find that this quantum dynamics implies a quantum extension of the 
Einstein equations.  The conserved energy-momentum tensor now contains a quantum 
contribution.  This conceptually-simple set up is promising both for deriving semiclassical and weak field approximations to the quantum Einstein equations and is thus important for the development of quantum corrections to computational general relativity.
\end{abstract}

\section{Introduction}
The problem of quantum gravity can be approached from many sides.  One of the 
major theoretical concerns however is to avoid an observer-based description as 
is the hallmark of e.g.\ the Copenhagen version of quantum mechanics.  The point 
is indeed that our theory must give a consistent physical cosmology, which does 
not suppose outside observers. Fundamental theories must directly address 
the beables of nature, and not the observables \cite{bell87a}. 

We formulate a theory which achieves this, involving geometry and matter as dynamical variables. More specifically, we develop a quantum modification of {\em geometrodynamics}, which describes the time-evolution of the 3-metric, just as in the Hamiltonian formulation of Einstein's gravity, coupled to nonrelativistic point particles. This is a Bohmian dynamics where the velocity of the 3-metric and the particle positions is determined by a wave function which itself satisfies a Schr\"odinger-type equation --- the Wheeler-DeWitt equation --- obtained from quantizing Einstein's gravity. From this quantum geometrodynamics, a modification to the Einstein equations is derived in the form of an extra contribution to the energy-momentum tensor. 

This extends the results of \cite{duerr20b} which applied to a matter field. Non-relativistic particles rather than relativistic particles are considered to make the analysis less technical. (The extension to relativistic particles could be done using e.g.\ \cite{struyve20b}.)

This work serves as a starting point to find quantum corrections to classical gravity, in particular applicable to the recently booming field of computational (classical) general relativity. It can be used to develop semi-classical approximations \cite{struyve17a,struyve20a} or weak field approximations, perhaps relevant for the study of of quantum modifications to gravitational waves. Another avenue is to investigate the status of space-time singularities in quantum gravity \cite{falciano15,demaerel19}. 

The outline of the paper is as follows. The next Section recalls the ADM formalism for classical gravity. It gives us the Hamiltonian formulation with constraints which can be quantized to lead to the appropriate Wheeler-DeWitt equation.  In Section \ref{bqm}, we obtain the velocity field that determines the quantum dynamics for particles positions and three-metric. From that dynamics we derive the quantum Einstein equations, with a conserved energy-momentum tensor containing a quantum contribution. We conclude with discussion and outlook in Section \ref{outl}.

\section{Geometrodynamics coupled to point-particles}
To formulate the classical theory, in particular the Hamiltonian form, the ADM formalism \cite{arnowitt62,kiefer04,bojowald11,carlip19} is employed. It is supposed that spacetime can be foliated in terms of space-like hypersurfaces such that the spacetime manifold is diffeomorphic to ${\mathbb R} \times \Sigma$, with $\Sigma$ a 3-surface. Coordinates $x^\mu=(t,{\bf x})$ can be chosen such that the time coordinate $t$ labels the leaves of the foliation and ${\bf x}$ are the coordinates on $\Sigma$.{\footnote{Units are chosen such that $\hbar=c=1$.}} In terms of these coordinates and following the conventions of \cite{carlip19}, the metric and its inverse are 
\be
g_{\mu \nu}=
\begin{pmatrix}
N^2 - N_i N^i & -N_i \\
-N_i & - h_{ij}
\end{pmatrix} \,,
\qquad
g^{\mu \nu}=
\begin{pmatrix}
\frac{1}{N^2} & \frac{-N^i}{N^2} \\
\frac{-N^i}{N^2} &   \frac{N^iN^j}{N^2}- h^{ij}
\end{pmatrix} \,,
\label{199}
\en
where $N$ and $N^i$ are respectively the lapse and the shift vector, and $h_{ij}$ is the Riemannian metric on $\Sigma$, with determinant $h$. Spatial indices are raised and lowered by this spatial metric. 

The total action is
\be
{\mathcal S}_{\textrm{cl}} =\int dt  \, L_{\textrm{cl}}=  \int d^4x \, 
{\mathcal L}_{\textrm{cl}}={\mathcal S}_G + {\mathcal S}_{M} ,
\label{203}
\en
which adds to the Einstein-Hilbert action ${\mathcal S}_G$ of gravity, with $\ka = 16\pi G$,
 \be
{\mathcal S}_G = -\frac{1}{\ka} \int d^4 x  \, \sqrt{-g}\, (R + 2 \Lambda) ,
\label{202}
\en
the action ${\mathcal S}_M$ for a non-relativistic particle in a gravitational field $g_{\mu \nu}$, using \eqref{199}, 
\begin{multline}
{\mathcal S}_{M} = - m \int dt N(t,{\bf X}(t))  \\
+ \frac{m}{2} \int dt \frac{1}{N(t, {\bf X}(t))} h_{ij}(t,{\bf X}(t)) \left[\dot 
X^i(t) + N^i(t, {\bf X}(t)) \right] \left[\dot X^j(t) + N^j(t,{\bf X}(t))\right] 
.
\label{201}
\end{multline}
The latter, with the first terms corresponding to the rest mass contribution, is derived as the non-relativistic limit of the relativistic particle action. As in classical electrodynamics \cite{spohn04}, a smearing could be introduced to avoid the singular nature of point-particles, but that is something we will not pursue here.

The action ${\mathcal S}_{\textrm{cl}}$ can also be expressed in terms of phase space variables. The canonically conjugate momenta are 
\be
P_i = \frac{\pa L_{\textrm{cl}} }{ \pa {\dot X}^i} = \frac{m}{N({\bf X})}\left[\dot X_i + N_i({\bf X}) \right] , \qquad \pi^{ij} = \frac{\pa {\mathcal L}_{\textrm{cl}} }{ \pa {\dot h}_{ij}} = - \frac{1}{\kappa} \sqrt{h} (K^{ij} - K h^{ij}),
\en
with 
\be
K_{ij} = \frac{1}{2N}({ D _i N_j + D _j N_i - \dot{h}}_{ij}) ,\qquad K = K_{ij}h^{ij},
\en
the extrinsic curvature. Indices of the particle position and momentum are raised and lowered with the metric $h_{ij}({\bf X})$. (The conjugate momentum is a tensor density of weight $-1$). In terms of these variables, the action reads
\be
{\mathcal S}_{\textrm{cl}} = \int dt \int_\Sigma d^3x  {\dot h}_{ij} \pi^{ij} + \int dt  {\dot X}^i P_i   - \int dt H,
\label{classicalaction}
\en
where $H$ is the Hamiltonian, given by
\be
H = \int_\Sigma d^3 x \left(N  {\mathcal H} + N^i  {\mathcal H}_i  \right),
\en
with
\be
{\mathcal H}  = {\mathcal H}_G + {\mathcal H}_M , \qquad {\mathcal H}_i  = {\mathcal H}_{Gi} + {\mathcal H}_{Mi}
\en
and
\be
 {\mathcal H}_G = \ka G_{ijkl} \pi^{ij} \pi^{kl} + {\mathcal V}, \qquad   {\mathcal H}_M({\bf x}) =   \delta({\bf x} - {\bf X}) \left( m+ \frac{P_iP^i}{2m}\right), 
\en
\be
 {\mathcal H}_{Gi} =  -2 h_{ik} D_j \pi^{jk}  ,\qquad   {\mathcal H}_{Mi}({\bf x}) = -\delta({\bf x} - {\bf X}) P_i  ,   
\en
where $G_{ijkl} = (h_{ij} h_{jl} + h_{il} h_{jk} -  h_{ij} h_{kl})/2\sqrt{h}$ is the DeWitt metric, $D_i$ is the covariant derivative corresponding to $h_{ij}$, and ${\mathcal V} = \sqrt{h}( 2\Lambda - R^{(3)})/\ka$ is the gravitational potential density. We write $V(t) = \int_\Sigma d^3 x  N(t,{\bf x}){\mathcal V}(h(t,{\bf x}))$.

In terms of these variables, the classical equations of motion are
\begin{align}
\dot{X}^i &= \frac{N({\bf X})}{m} P^i - N^i(\textbf{X}),\label{c11bis}\\
\dot P_i &= - \pa_i N \left(m + \frac{1}{2m} P^kP_k\right) - \frac{N}{2m} \pa_i 
h^{kl} P_k P_l + \pa_i N^k P_k,
\label{c12}\\
{\dot{h}}_{ij} &= 2\ka NG_{ijkl}\pi^{kl} + D _i N_j + D _j N_i,
\label{c10}\\
{\dot{\pi}}^{ij}({\bf x}) &= - N({\bf x}) \ka \frac{\pa G_{mnkl}}{\pa 
h_{ij}({\bf x})} \pi^{mn}({\bf x}) \pi^{kl}({\bf x}) \nonumber\\
&\phantom{{}={}} -  \frac{\delta V}{ \de h_{ij}({\bf x})}  - \frac{\de }{ \de 
h_{ij}({\bf x})} \int_\Sigma d^3y \left(-2 N_k D_l \pi^{kl} \right) + 
\delta({\bf x} - {\bf X})\frac{N({\bf X})}{2m} P^i P^j,
\label{c11}\\
{\mathcal H}  &= 0,\qquad {\mathcal H}_i  = 0  .
\label{c13}
\end{align}
The last two equations are the Hamiltonian and diffeomorphism constraint. While equation \eqref{c11} can be worked out more, that is not necessary for our purposes.  For the full expression, see e.g.\ \cite{wald84,bojowald11}.\\

Together, Equations \eqref{c10}--\eqref{c13} constitute the Einstein field equations
\be
G_{\mu \nu } - \Lambda g_{\mu \nu}= \frac{\kappa}{2} T_{\mu \nu}, 
\label{c20}
\en
where $G_{\mu \nu } = R_{\mu \nu } - \frac{1}{2}Rg_{\mu \nu} $ is the Einstein tensor and is the energy-momentum tensor, with components
\begin{align}
T^{00}(t,{\bf x}) &= \frac{1}{N(t,{\bf x})^2\sqrt{h(t,{\bf x})}} \left[ m + \frac{1}{2m}P_k(t) P^k(t)\right] \delta({\bf x} - {\bf X}(t)) , \nonumber\\
T^{0i}(t,{\bf x})&=T^{i0}(t,{\bf x}) = \frac{P^i(t)}{N(t,{\bf x}){\sqrt{h(t,{\bf x})}}}\delta({\bf x} - {\bf X}(t)) - N^i(t,{\bf x})T^{00}(t,{\bf x}), \nonumber\\
T^{ij}(t,{\bf x}) &= \frac{1}{{\sqrt{h(t,{\bf x})}}}\left[ \frac{P^i(t)P^j(t)}{m} -\frac{1}{N(t,{\bf x})} \big( P^iN^j(t,{\bf x}) + P^jN^i(t,{\bf x})\big)\right]\delta({\bf x} - {\bf X}(t)) \nonumber\\
& \qquad + N^i(t,{\bf x})N^j(t,{\bf x})T^{00}(t,{\bf x}) , 
\label{emtensor}
\end{align}
which is most easily derived using $T^{\mu \nu} =  - \frac{2}{\sqrt{-g}} \frac{\de {\mathcal S}_{M}}{\de g_{\mu \nu}}$ and the relations
\begin{align*}
\frac{\pa}{\pa g_{00}} &= \frac{1}{2N} \frac{\pa}{\pa N} ,\quad  \frac{\pa}{\pa 
g_{0i}} = - \frac{N^i}{2N} \frac{\pa}{\pa N} - \frac{h^{ik}}{2}\frac{\pa}{\pa 
N^k},\\ \frac{\pa}{\pa g_{ij}} &= - \frac{\pa}{\pa h_{ij}} - \frac{N^iN^j}{2N} 
\frac{\pa}{\pa N} +  \frac{N^{i} h^{jk} + N^j h^{ik}}{2} \frac{\pa}{\pa N^k}.
\end{align*}

\section{Wave mechanics for 3-metric and particles}\label{bqm}
The ADM formalism of Einstein gravity is the starting point for canonical quantization~\cite{kiefer04}. In the present case, this results in a wave functional $\Psi(\mathbf{X},h)$ which is a function of the 3-metric $h$ and the particle coordinates ${\bf X}$ which satisfies the Wheeler-DeWitt equation
\be
{\widehat {\mathcal H}} \Psi \equiv {\widehat {\mathcal H}}_G \Psi + {\widehat {\mathcal H}}_M \Psi= 0,
\label{q1}
\en
with diffeomorphism constraint
\be
{\widehat {\mathcal H}}_i \Psi \equiv {\widehat {\mathcal H}}_{Gi} \Psi + {\widehat {\mathcal H}}_{Mi} \Psi= 0,
\label{q2}
\en
with{\footnote{We have chosen the Laplace-Beltrami ordering for matter, but not 
for gravity. Other choices are of course possible 
\cite{kuchar73,christodoulakis86,duerr20b}.}}
\be
{\widehat {\mathcal H}}_G   =  - \ka G_{ijkl} \frac{\delta}{\delta h_{ij}} 
\frac{\delta}{\delta h_{kl}}  + {\mathcal V}(h,\phi), \qquad {\widehat 
{\mathcal H}}_M({\bf x}) =   \delta({\bf x} - {\bf X}) \left( m - 
\frac{\nabla^2 }{2m}\right),
\label{q3}
\en
\be
{\widehat {\mathcal H}}_{Gi}  =   2 h_{ik}D_j\frac{\delta }{\delta h_{jk}}, 
\qquad  {\widehat {\mathcal H}}_{Mi}({\bf x}) =  \delta({\bf x} - {\bf X}) \nabla_i ,
\en
where $\nabla_i$ corresponds to the covariant derivative $D_i$ evaluated at ${\bf X}$, with $\nabla_i \psi = \pa_i \psi$, and $\nabla^2 = \na_i\na^i$. \\

Writing $\Psi=|\Psi|\ee^{\ii S}$, we define the guidance equations for $(\mathbf{X},h)$ as
\begin{align}
\dot X^i &=   \frac{N({\bf X})}{m} \nabla^i S(h,{\bf X}) - N^i({\bf X}),\label{dotphi}\\
{\dot{h}}_{ij} &= 2\ka NG_{ijkl}\frac{\delta S}{\delta h_{kl}} + D _i N_j + D_j N_i.
\label{bcqg1}
\end{align}
This amounts to a Bohmian dynamics, developed in other contexts in \cite{bohm93,holland93b,duerr09,pinto-neto19}. As in the classical case, the dynamics involves a lapse and shift function. Together with the 3-metric they determine a 
4-metric defined by \eqref{199}. Classically, the choice of lapse and shift do 
not change the 4-geometry (i.e.\ the coordinate independent content of the 
metric). However, this is no longer the case for the quantum dynamics; 
different choices of the shift vector give the same geometry, but lapse 
functions do not (unless they merely differ by a factor which depends only on 
time). Put differently, the dynamics depends on the particular choice 
of foliation of spacetime \cite{pinto-neto19}. That does not necessarily imply 
the predictions of the quantum theory depend on that choice. (The Bohmian dynamics in Minkowski space-time may also violate Lorentz invariance at the fundamental level, yet the stastistical predictions are Lorentz invariant \cite{duerr14}.)

\section{Hamiltonian form of the quantum dynamics}
To derive the quantum Einstein equations, we first derive a Hamiltonian form of the dynamics. Defining 
\be
\label{q15}
P_i = \nabla_i S , \qquad \pi^{ij}=\frac{\delta S}{\delta h_{ij}},
\en
the quantum dynamics \eqref{dotphi}--\eqref{bcqg1} obtains the form 
\be
\dot X^i =   \frac{N({\bf X})}{m} P^i  - N^i({\bf X}), \quad {\dot{h}}_{ij} = 2\ka NG_{ijkl}\pi^{kl} + D _i N_j + D _j N_i,
\label{bcqg1-2}
\en
which agrees with the classical dynamics \eqref{c11bis} and \eqref{c10}.

The constraint \eqref{q1} implies that
\be
 \ka G_{ijkl}({\bf x})\frac{\delta S}{\delta h_{ij}({\bf x})} \frac{\delta S}{ \delta h_{kl}({\bf x})} + {\mathcal V}({\bf x}) + \frac{\nabla_i S \na^i S}{2m}\delta({\bf x} - {\bf X}) + {\mathcal Q}({\bf x})=0,
\label{q10}
\en
where ${\mathcal Q}$ is the quantum potential density, given by
\be\label{qptot}
{\mathcal Q} = {\mathcal Q}_{ G} + {\mathcal Q}_{M},
\en
with
\be
{\mathcal Q}_{ G} = -\ka G_{ijkl} \frac{1}{|\Psi|}\frac{\delta|\Psi|}{\delta h_{ij}\delta h_{kl}}  , \qquad {\mathcal Q}_{M}({\bf x}) =- \delta({\bf x} - {\bf X}) \frac{\na^2 |\Psi|}{2m|\Psi|} .
\en
From the other constraint \eqref{q2} we find that
\be
2 D_j\frac{\delta S}{\delta h_{ji}({\bf x})} + \delta({\bf x} - {\bf X})\nabla^i S  = 0.
\label{q11}
\en
Hence, using the definition \eqref{q15}, we have that
\be
{\mathcal H} + {\mathcal Q} = 0, \qquad {\mathcal H}_i = 0.
\label{h2}
\en
These are the classical constraint equations \eqref{c13} modified by ${\mathcal V} \to {\mathcal V} + {\mathcal Q}$. 

Finally, taking the time-derivative of \eqref{q15} and using \eqref{q10}, 
\eqref{q11} and defining $Q(t) = \int_\Sigma d^3 x  N({\bf x},t){\mathcal 
Q}({\bf x},t)$, it follows that
\begin{align}
\dot P_i =& - \pa_i N \left(m + \frac{1}{2m} P^kP_k\right) - \frac{N}{2m} \pa_i 
h^{kl} P_k P_l + \pa_i N^k P_k - \pa_i Q,\nonumber\\
{\dot{\pi}}^{ij}({\bf x}) =& - N({\bf x}) \ka \frac{\pa G_{mnkl}}{\pa 
h_{ij}({\bf x})} \pi^{mn}({\bf x}) \pi^{kl}({\bf x}) \\
&-  \frac{\delta }{ \de h_{ij}({\bf x})}(V+Q)  - \frac{\de }{ \de h_{ij}({\bf x})} \int_\Sigma d^3y \left(-2 N_k D_l \pi^{kl} \right) + \delta({\bf x} - {\bf X})\frac{N({\bf X})}{2m} P^i P^j ,
\label{h4}
\end{align}
which are the classical equations \eqref{c12}, \eqref{c11}, modified by $V \to  V + Q$. In this way, a Hamiltonian form of the dynamics is obtained, given by \eqref{bcqg1-2}, \eqref{h2}-\eqref{h4}.\\

A crucial advantage of this Hamiltonian formulation is that it can be derived from the action
\be
{\mathcal S} = {\mathcal S}_{\textrm{cl}} + {\mathcal S}_{Q},
\label{baction}
\en
with 
\be
{\mathcal S}_Q = - \int dt\,  Q = - \int dt \int_\Sigma d^3x \,N\,{\mathcal Q}.
\en
%
The Bohmian dynamics is not completely determined by this action as it also entails \eqref{q15} which should be imposed here as an additional constraint. Actually, it is sufficient to assume that \eqref{q15} holds on a certain leaf of the foliation \cite{pinto-neto99}, because then it will hold on all leaves of the foliation.\\  
From \eqref{baction} we now immediately get the quantum Einstein equations by varying with respect to $g_{\mu \nu}$:
\be
G_{\mu \nu } - \Lambda g_{\mu \nu}= \frac{\kappa}{2} \left( T_{\mu \nu} + T_{Q \mu \nu}\right),
\label{qee}
\en
with $T^{ \mu \nu}$ the classical energy-momentum tensor given in \eqref{emtensor} and 
\[
T^{ \mu \nu}_Q= -\frac{2}{\sqrt{-g(x)}} \frac{\de {\mathcal S}_Q}{\de g_{\mu \nu}(x)}
\]
 a quantum contribution given by
\begin{align}
T^{00}_Q(t,{\bf x}) &= \frac{1}{N(t,{\bf x})^2\sqrt{h(t,{\bf x})}} {\mathcal Q}(t,{\bf x}) , \nonumber\\
T^{0i}_Q(t,{\bf x})&=T^{i0}_Q(t,{\bf x}) =  - N^i(t,{\bf x})T^{00}_Q(t,{\bf x}) , \nonumber\\
T^{ij}_Q(t,{\bf x}) &=  N^i(t,{\bf x})N^j(t,{\bf x})T^{00}_Q(t,{\bf x}) -  \frac{2}{N({\bf x},t)\sqrt{h({\bf x},t)}}\frac{\de }{\de h_{ij}({\bf x})}\int_\Sigma d^3y N({\bf y},t){\mathcal Q}({\bf y},t). \label{quantumtensor}
\end{align}
This tensor has the same form as in the scalar field case of \cite{duerr20b}, but with a different quantum potential. Defining ${\widetilde {\mathcal Q}} = {\mathcal Q}/\sqrt{h}$, it can be written more compactly as
\be
T_{Q}^{\mu \nu}(x) = g^{\mu \nu}(x) {\widetilde {\mathcal Q}}(x) - \frac{2}{\sqrt{-g(x)}} \de^\mu_i \de^\nu_j \int_\Sigma d^3y {\sqrt{-g({\bf y},t)}}   \frac{\de {\widetilde {\mathcal Q}}({\bf y},t)}{\de h_{ij}({\bf x})}  .
\label{quantumtensor2}
\en
Note that if ${\widetilde {\mathcal Q}}$ is constant, then it acts as a cosmological constant. This feature has been explored in \cite{squires92,pinto-neto03,demaerel20}.\\


So far, we have just considered a single particle. The generalization to many particles is straightforward. The wave functional now depends on all the particle positions. i.e., $\Psi=\Psi({\bf X}_1,\dots,{\bf X}_n,h)$ and satisfies the constraints \eqref{q1} and \eqref{q2}, with
\be
{\widehat {\mathcal H}}_M({\bf x}) =  \sum_k N({\bf x}) \delta({\bf x} - {\bf X}_k) \left( m_k - \frac{\nabla^2_k }{2m_k}\right), \qquad  {\widehat {\mathcal H}}_{Mi}({\bf x}) =  \sum_k \delta({\bf x} - {\bf X}_k) \nabla_{ki} ,
\label{q3many}
\en
where $k$ is the particle label. The dynamics for the metric remains \eqref{bcqg1}, while that of the particles is
\be
\dot X^i_k =   \frac{N({\bf X}_k)}{m_k} \nabla^i S({\bf X}_1,\dots,{\bf X}_n,h) - N^i({\bf X}_k).
\label{dotphimany}
\en
The form of the quantum Einstein equations also remains unchanged, with the proviso that the matter part of the quantum potential is now
\be
{\mathcal Q}_{M}({\bf x}) =- \sum_k \delta({\bf x} - {\bf X}_k) \frac{\na^2_k |\Psi|}{2m_k|\Psi|} .
\en

\section{Discussion and outlook}\label{outl}
A direct consequence of the quantum Einstein equations \eqref{qee} is that the energy-momentum tensor $T_{\mu \nu} + T_{Q \mu \nu}$ is covariantly conserved. Namely, the left-hand-side of \eqref{qee} is covariantly conserved as an identity. Consistency implies that the right-hand-side is also covariantly conserved. As such, the status of energy conservation is the same as in classical general relativity \cite{duerr20b}. This is in contrast with non-relativistic Bohmian mechanics (in a given Euclidean or Riemannian metric), where the total energy is not conserved if the quantum potential is explicitly time-dependent.\\

This work will be of interest to develop a semi-classical or weak field approximations. Namely, one goal would be to derive an approximation where gravity is treated classically and the particles quantum mechanically. Our energy-momentum tensor would then act as the source in the classical Einstein field equations. As explained in detail in \cite{struyve20a}, care has to be taken of the gauge invariance (in this case the invariance under spatial diffeomorphisms), by separating gauge degrees of freedom from gauge invariant degrees of freedom. This is easily done in the case of electromagnetism using for example the transverse degrees of freedom of the electric field \cite{struyve20a}. Presumably something similar can be done in the case of gravity, by passing to the weak field limit and considering the transverse traceless degrees of freedom, perhaps following the ideas of \cite{anastapopoulos13}.

\section{Acknowledgments}
WS is supported by the Research Foundation Flanders (Fonds Wetenschappelijk Onderzoek, FWO), Grant No. G066918N. 

\end{document}